\documentclass[a4paper,usenames,dvipsnames,11pt]{article}
\pdfoutput=1

\usepackage{jheppub}
\usepackage{slashed}
\usepackage{mathrsfs,booktabs,multirow,tabularx}
\usepackage{stmaryrd}
\usepackage{xspace}
\usepackage{fancyvrb}
\usepackage[makeroom]{cancel}

\def\t{{\bar t}}
\def\Mtt{m(t\bar t)}

\def\PTt{p_{T,t}}
\def\PTtbar{p_{T,{\bar t}}}
\def\PTavt{p_{T,{\rm avt}}}
\def\Yavt{y_{\rm avt}}
\def\Ytt{y(t\bar t)}
\def\GeV{\, \rm GeV}
\def\TeV{\, \rm TeV}

\def\alphas{\alpha_s}
\def\SigmaSub{\Sigma_{\rm res}}

\title{Top-pair production at the LHC through NNLO QCD and NLO EW}

\author[a]{Micha\l{}  Czakon,}
\author[b]{David Heymes,}
\author[b]{Alexander Mitov,}
\author[c,d]{Davide Pagani,}
\author[c]{Ioannis Tsinikos}
\author[e,f]{and Marco Zaro}

\affiliation[a]{Institut f\"ur Theoretische Teilchenphysik und Kosmologie,
RWTH Aachen University, Aachen, D-52056 Germany}
\affiliation[b]{Cavendish Laboratory, University of Cambridge, Cambridge CB3 0HE, UK}
\affiliation[c]{Centre for Cosmology, Particle Physics and Phenomenology (CP3), Universit\'e Catholique de Louvain, Chemin du Cyclotron 2, Louvain la Neuve, B-1348 Belgium}
\affiliation[d]{Technische Universit\"{a}t M\"{u}nchen, James-Franck-Str. 1, Garching, D-85748 Germany}
\affiliation[e]{Sorbonne Universit\'es, UPMC Univ. Paris 06, UMR 7589, LPTHE, Paris, F-75005 France}
\affiliation[f]{CNRS, UMR 7589, LPTHE, Paris, F-75005 France}

\emailAdd{mczakon@physik.rwth-aachen.de}
\emailAdd{dheymes@hep.phy.cam.ac.uk}
\emailAdd{adm74@cam.ac.uk}
\emailAdd{davide.pagani@tum.de}
\emailAdd{ioannis.tsinikos@uclouvain.be}
\emailAdd{marco.zaro@lpthe.jussieu.fr}

\note{Preprint: Cavendish-HEP-17/07, CP3-17-12, TUM-HEP-1084/17, TTK-17-15}

\abstract{In this work we present for the first time predictions for top-quark pair differential distributions at the LHC at NNLO QCD accuracy and including EW corrections. For the latter we include not only contributions of ${\cal O}(\alphas^2 \alpha)$, but also those of order ${\cal O}(\alphas \alpha^2)$ and ${\cal O}( \alpha^3)$.  Besides providing phenomenological predictions for all main differential distributions with stable top quarks, we also study the following issues. 1) The effect of the photon PDF on top-pair spectra: we find it to be strongly dependent on the PDF set used -- especially for the top $p_T$ distribution. 2) The difference between the additive and multiplicative approaches for combining QCD and EW corrections: with our scale choice, we find relatively small differences between the central predictions, but reduced scale dependence within the multiplicative approach. 3) The potential effect from the radiation of heavy bosons on inclusive top-pair spectra: we find it to be, typically, negligible.}

\begin{document} 
\maketitle
\flushbottom

\section{Introduction}\label{sec:intro}

The availability of NNLO QCD predictions for stable top-pair production at the LHC, both for the total cross-section \cite{Baernreuther:2012ws,Czakon:2012zr,Czakon:2012pz,Czakon:2013goa} with NNLL soft-gluon resummation \cite{Cacciari:2011hy,Czakon:2011xx} and for all the main differential distributions \cite{Czakon:2015owf,Czakon:2016dgf}, has made it possible to compare Standard Model (SM) theory with LHC data at the few-percent level accuracy. Such a high precision has led, among others, to further scrutiny of the differences between LHC measurements \cite{Czakon:2016olj} and the ability of Monte-Carlo event generators to describe hadronic $t\t$ production. As a result of these ongoing studies, new MC developments are taking place, such as the incorporation of non-resonant and interference effects \cite{Jezo:2016ujg,Frederix:2016rdc}, which builds upon previous works that included NLO top decay corrections through-fixed order \cite{Melnikov:2009dn,Bevilacqua:2010qb,Denner:2012yc,Bevilacqua:2015qha,Bevilacqua:2016jfk} and/or showered \cite{Norrbin:2000uu,Campbell:2014kua,Jezo:2015aia} calculations.

One of the remaining ways for further improving SM theory predictions is by consistently including the so called Electro-Weak (EW) corrections on top of the NNLO QCD ones. Weak ~\cite{Beenakker:1993yr,Kuhn:2005it, Bernreuther:2005is, Kuhn:2006vh,Bernreuther:2006vg,Bernreuther:2008md, Manohar:2012rs, Kuhn:2013zoa,Campbell:2015vua}, QED~\cite{Hollik:2007sw} and EW (weak+QED) \cite{Bernreuther:2010ny, Hollik:2011ps, Kuhn:2011ri, Bernreuther:2012sx, Pagani:2016caq} corrections to top-quark pair production have been known for quite some time, and also EW corrections to the fully off-shell dilepton signature are nowadays available \cite{Denner:2016jyo}. As it has been documented in the literature, although EW effects are rather small at the level of total cross-section, they can have a sizeable impact on differential distributions and also on the top-quark charge asymmetry.

The goal of this work is to consistently merge existing NNLO QCD predictions with EW corrections  into a single coherent prediction and to study its phenomenological impact. 
This is achieved by combining the NNLO QCD predictions from ref.~\cite{Czakon:2016dgf} with the complete LO and NLO contributions derived within the framework of ref.~\cite{Pagani:2016caq}. Specifically, we include the NLO EW effects of ${\cal O}(\alphas^2 \alpha)$, all subleading NLO (${\cal O}(\alphas \alpha^2)$ and ${\cal O}( \alpha^3)$) terms as well as the LO (${\cal O}(\alphas \alpha)$ and ${\cal O}( \alpha^2)$) contributions.

Another motivation for this study stems from recent developments in understanding the photon content of the proton \cite{Harland-Lang:2016kog, Manohar:2016nzj}. As shown in ref.~\cite{Pagani:2016caq}, depending on the PDF set, photon-initiated contributions can be numerically significant in some regions of phase space.~\footnote{This has been studied also in refs.~\cite{Accomando:2016tah,Accomando:2016ehi} for the case of neutral-current Drell-Yan production.} If the photon density from the {\sc\small NNPDF3.0QED} set \cite{Ball:2013hta,Ball:2014uwa} is employed, the photon-initiated contribution is large in size and of opposite sign with respect to the Sudakov EW corrections, leading to the almost complete cancellation of the two effects. Nevertheless, large PDF uncertainties from the photon PDF are still present after this cancellation. On the other hand, theoretical consensus about the correctness of the novel approach introduced in ref.~\cite{Manohar:2016nzj} appears to have emerged by now.
\footnote{The consensus has been also supported by a preliminary study in the determination of the photon PDF including new LHC data \cite{Aad:2016zzw}.}
The PDF set provided with ref.~\cite{Manohar:2016nzj}, named {\sc\small LUXQED}, includes a photon PDF whose central value and relative uncertainty are both much smaller than in the case of {\sc\small NNPDF3.0QED}. Thus, at variance with the {\sc\small NNPDF3.0QED} set, neither large cancellation between Sudakov effects and photon-induced contributions nor large photon PDF uncertainty is present in {\sc\small LUXQED}-based predictions.  

In order to document the ambiguity arising from the differences between the photon densities in the available PDF sets, with the exception of sec.~\ref{sec:pheno}, in this work we always give predictions for top-pair differential distributions at the LHC based on the {\sc\small LUXQED} \cite{Manohar:2016nzj} and {\sc\small NNPDF3.0QED} \cite{Ball:2013hta,Ball:2014uwa} PDF sets. We believe that our findings will provide a valuable input to future PDF determinations including EW effects.

This paper is organised as follows: sec.~\ref{sec:pheno} is devoted to the phenomenological study of our combined QCD and EW predictions for the LHC at 13 TeV. The reasons behind some of the choices made in sec.~\ref{sec:pheno} -- like the choice of PDF set and combination approach -- are revealed in sec.~\ref{sec:compare}, where we compare in great detail two approaches for combining NNLO QCD and EW corrections in top-pair differential distributions. The so-called additive approach is discussed in sec.~\ref{sec:main}, while the multiplicative one in sec.~\ref{sec:multiplic}. Section~\ref{sec:photonpdf} is dedicated to studying the impact of the photon PDF on top-pair spectra.  In sec.~\ref{sec:HBR} we provide an estimate of the impact of inclusive Heavy Boson Radiation (HBR), namely the contribution from $t\t V$ final states with $V=H,W^{\pm},Z$. While most of the notation is introduced in the main text some technical details are delegated to Appendix \ref{sec:appendix}.

\section{Phenomenological predictions for the LHC at 13 TeV}\label{sec:pheno}

In this section we  present predictions for $t\t$ distributions for the LHC at 13 TeV at NNLO QCD accuracy including also EW corrections. We focus on the following distributions: the top-pair invariant mass $m(t\t)$, the top/antitop average transverse momentum ($\PTavt$) and rapidity ($\Yavt$) and the rapidity $y(t\t)$ of the $t\t$ system. The $\PTavt$ ($\Yavt$) distributions are calculated {\it not} on an event-by-event basis but by averaging the results of the histograms for the transverse momentum (rapidity) of the top and the antitop.

Our calculation is performed using the following input parameters
\begin{equation}
m_t = 173.3 \text{ GeV}\, , \quad m_H = 125.09 \text{ GeV} \, , \quad m_W = 80.385 \text{ GeV} \, , \quad m_Z = 91.1876 \text{ GeV} \,,
\end{equation}
while all other fermion masses are set to zero. All masses are renormalised on-shell and all decay widths are set to zero. The renormalisation of $\alphas$ is performed in the 5-flavour scheme while EW input parameters and the associated $\alpha$ renormalisation condition are in the $G_\mu$-scheme, with
\begin{equation}
G_\mu = 1.1663787 \cdot 10^{-5} \text{ GeV}^{-2} \,.
\end{equation}

The EW corrections have been calculated in a completely automated way via an extension of the {\sc\small MadGraph5\_aMC@NLO} code \cite{Alwall:2014hca} that has been already validated in refs.~\cite{Frixione:2015zaa, Badger:2016bpw}, and in ref.~\cite{Frederix:2016ost} for the calculation of the complete NLO corrections. 

We work with dynamical renormalisation ($\mu_r$) and factorisation ($\mu_f$) scales. Their common central value is defined as
\begin{eqnarray}
\mu &=& \frac{m_{T,t}}{2}~~{\rm for~the} ~ \PTt ~ {\rm distribution}, \label{eq:scalemT}\\
\mu &=& \frac{m_{T,\t}}{2}~~{\rm for~the} ~ \PTtbar ~ {\rm distribution}, \label{eq:scalemTbar}\\
\mu &=& \frac{H_T}{4} = \frac{1}{4} \left( m_{T,t} + m_{T,\bar t} \right)~~{\rm for~all~other~distributions},
\label{eq:scaleHT}
\end{eqnarray}
where $m_{T,t}\equiv\sqrt{m_t^2+\PTt^2}$ and $m_{T,\bar t} \equiv\sqrt{m_t^2+\PTtbar^2}$ are the transverse masses of the top and antitop quarks. As already mentioned, $\PTavt$ and $\Yavt$ distributions are obtained by averaging the top and antitop distributions for the transverse momentum and rapidity, respectively.

These scale choices have been motivated and studied at length in ref.~\cite{Czakon:2016dgf}. In all cases theoretical uncertainties due to missing higher orders are estimated via the 7-point variation of $\mu_r$ and $\mu_f$ in the interval $\{\mu/2,2\mu\}$ with $1/2\leq\mu_r/\mu_f\leq2$. We remark that the combination of QCD and EW corrections is independently performed for each value of $\mu_{f,r}$.

For theoretical consistency, a set of PDFs including QED effects in the DGLAP evolution should always be preferred whenever NLO EW corrections are computed. At the moment, the only two  NNLO QCD accurate PDF sets that include them are {\sc\small NNPDF3.0QED} and {\sc\small LUXQED}.
~\footnote{The PDF sets {\sc\small  MRST2004QED} \cite{Martin:2004dh} and {\sc\small CT14QED} \cite{Schmidt:2015zda} also include QED effects in the DGLAP evolution, but they are not NNLO QCD accurate. A PDF set including full SM LO evolution (not only QCD and QED but also weak effects) has also recently become available \cite{Bauer:2017isx}.} 
Both sets have a photon density, which induces additional contributions to $t\t$ production \cite{Hollik:2007sw,Pagani:2016caq}.

As motivated and discussed at length in sec.~\ref{sec:compare}, the phenomenological predictions in this section are based on the {\sc\small LUXQED} PDF set and on the multiplicative approach for combining QCD and EW corrections, which we will denote as ${\rm QCD\times EW}$. We invite the interested reader to consult secs.~\ref{sec:compare} and \ref{sec:photonpdf} where detailed comparisons between the two PDF sets as well as between the two approaches for combining QCD and EW corrections can be found.

\begin{figure}[t]
\centering
\includegraphics[width=0.49\textwidth]{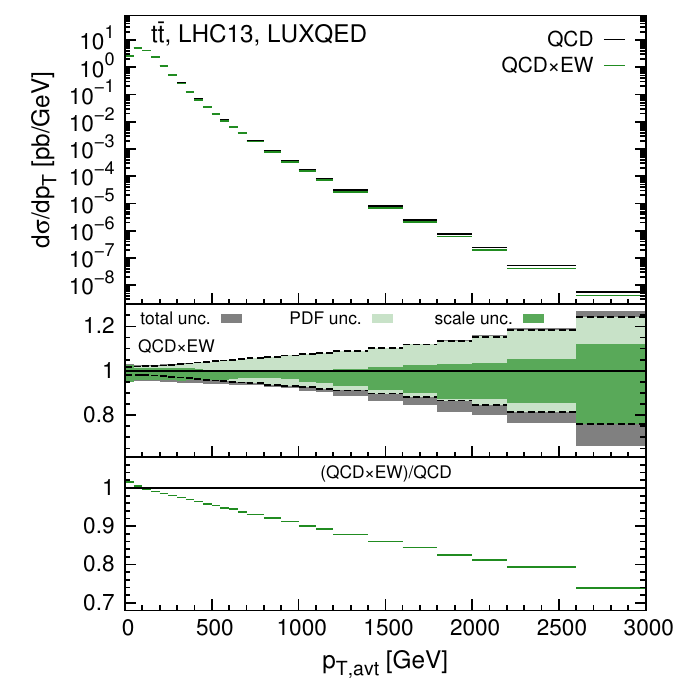}
\includegraphics[width=0.49\textwidth]{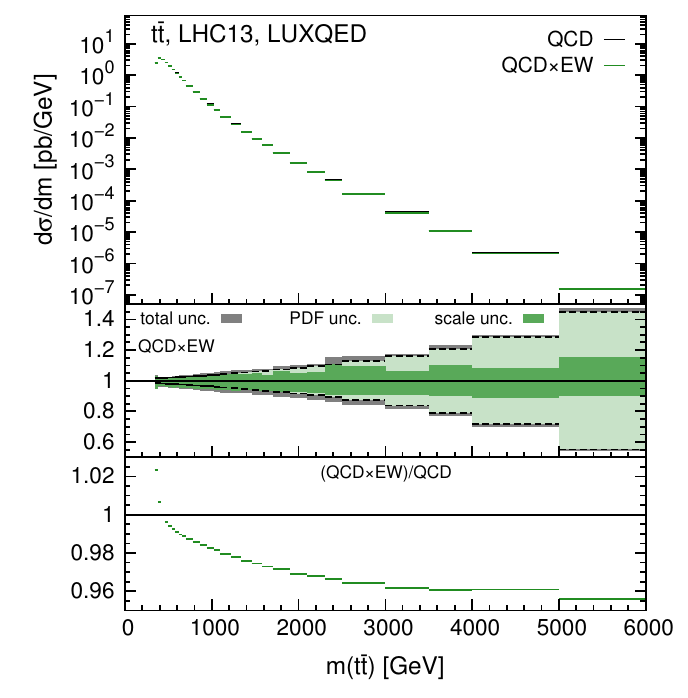}
\includegraphics[width=0.49\textwidth]{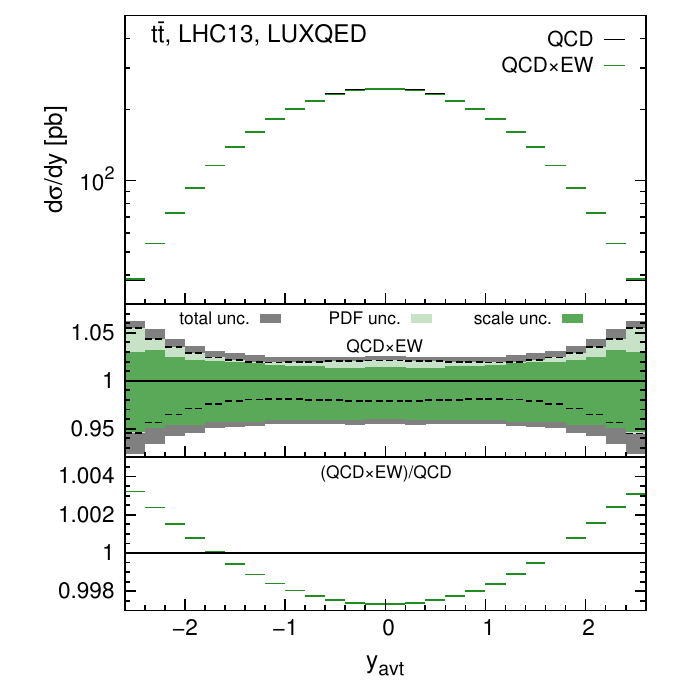}
\includegraphics[width=0.49\textwidth]{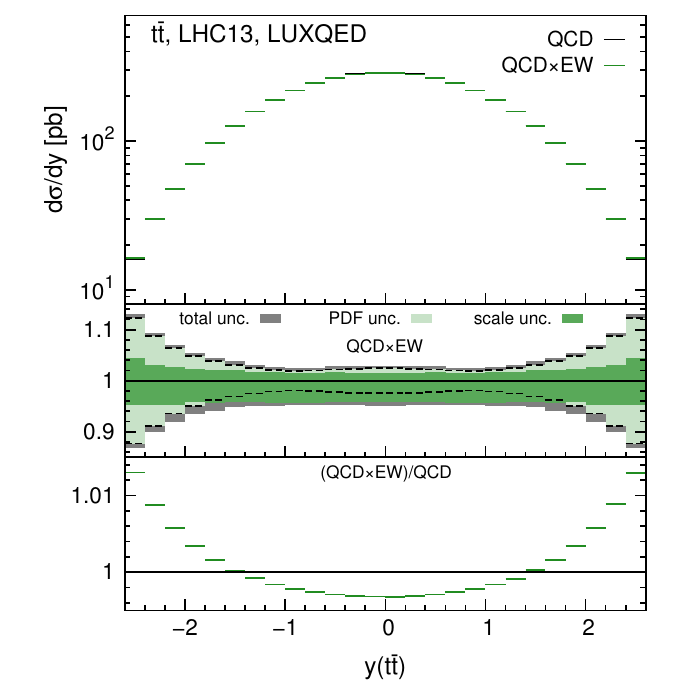}
\caption{Our best predictions for the four LHC 13 TeV $t\t$ differential distributions considered in this work. The predictions are based on the multiplicative approach. Shown are the scale, PDF and combined (in quadrature) theory uncertainties for each  ${\rm QCD\times EW}$ distribution. The boundaries of the PDF variation band are marked with black dashed lines. Also shown is the ratio of central scales for the combined QCD and EW prediction with respect to the NNLO QCD one.}
\label{fig:pheno}
\end{figure}
From the plots shown in fig.~\ref{fig:pheno} we conclude that the impact of the EW corrections relative to NNLO QCD depends strongly on the kinematic distribution. The smallest impact is observed in the two rapidity distributions: the relative effect for $\Yavt$ is around 2 permil and is much smaller than the scale uncertainty. The $\Ytt$ distribution is slightly more sensitive, with a relative impact of slightly above 1\% for large values of $\Ytt$. This correction is also well within the scale-variation uncertainty band. The impact of EW corrections on the $\Mtt$ distribution is larger. Relative to NNLO QCD it varies between +2\% at the absolute threshold and -6\% at high energies. Still, this correction is well within the scale variation uncertainty. The small sensitivity of the $\Yavt$ and $\Ytt$ distributions to EW corrections supports the findings of ref.~\cite{Czakon:2016olj} where these two distributions were used for constraining the PDF of the gluon. 

The largest correction due to EW effects is observed in the $\PTavt$ distribution. Relative to NNLO QCD, the correction ranges from +2\% at low $\PTavt$ to -25\% at $\PTavt\sim 3\TeV$. The correction is significant and is comparable to the scale variation band already for $\PTavt\sim 500\GeV$. Overall, the EW contribution to the $\PTavt$ distribution is as large as the total theory uncertainty band in the full kinematic range $\PTavt\leq 3\TeV$ considered in this work. 

The fraction of the theory uncertainty induced by PDFs is strongly dependent on kinematics.
For the $\Yavt$ and $\Ytt$ distributions, the PDF error is slightly smaller than the scale uncertainty for central rapidities, but is larger in the peripheral region, especially for the $\Ytt$ distribution.
The PDF uncertainty becomes the dominant source of theory error in the $\PTavt$ distribution for $\PTavt$ as large as $500\GeV$, while for the $\Mtt$ distribution it begins to dominate over the scale uncertainty for $\Mtt\sim 2.5\TeV$.

There are many applications for the results derived in this work. Examples are: inclusion of EW effects in PDF determinations from LHC $t\t$ distributions, high-mass LHC searches, precision SM LHC measurements and benchmarking of LHC event generators. A practical and sufficiently accurate procedure for the utilisation of our results could be as follows. One starts by deriving an analytic fit for the ${\rm QCD\times EW}/{\rm QCD}$ $K$-factor (all $K$-factors in fig.~\ref{fig:pheno} are available in electronic form \footnote{\label{Rep}Repository with results and additional plots of NNLO QCD + EW $t \bar t$ differential distributions: \url{http://www.precision.hep.phy.cam.ac.uk/results/ttbar-nnloqcd-nloew/}}); as evident from fig.~\ref{fig:pheno} it is a very smooth function for all four differential distributions. Under the assumption that this $K$-factor is PDF independent, such an analytic fit could then be used to rescale the NNLO--QCD--accurate differential distributions derived with any PDF set from existing NNLO QCD \cite{Czakon:2017dip}  {\sc fastNLO} \cite{Kluge:2006xs,Britzger:2012bs} tables. Regarding the PDF error of NNLO QCD differential distributions, it can be calculated very fast with any PDF set with the help of the {\sc fastNLO} tables of ref.~\cite{Czakon:2017dip}. As we show in the following the PDF error of the QCD and combined QCD and EW predictions is almost the same, especially for the {\sc\small LUXQED} PDF set used for our phenomenological predictions.

\section{Comparison of two approaches for combining NNLO QCD predictions and EW corrections}\label{sec:compare}

In this work we compare two approaches for combining QCD and EW corrections. For brevity, we will refer to them as {\it additive} and {\it multiplicative} approaches. As already mentioned, the results presented in sec.~\ref{sec:pheno} have been calculated using the multiplicative approach.

In the additive approach the NNLO QCD predictions (defined as the complete set of ${\cal O}(\alphas^n)$ terms up to $n=4$) are combined with all possible remaining LO and NLO terms arising from QCD and electroweak interactions in the Standard Model. In other words, at LO we include not only the purely QCD ${\cal O}(\alphas^2)$ contribution, but also all ${\cal O}(\alphas \alpha)$ and ${\cal O}(\alpha^2)$ terms. Similarly, at NLO we take into account not only the NLO QCD ${\cal O}(\alphas^3)$ contribution but also the ${\cal O}(\alphas^2 \alpha)$ one, the so-called NLO EW, as well as the subleading contributions of ${\cal O}(\alphas \alpha^2)$ and ${\cal O}( \alpha^3)$. For brevity, we will denote as ``EW corrections'' the sum of all LO and NLO terms of the form ${\cal O}(\alphas^{m}\alpha^{n})$ with $n>0$. Moreover, when we will refer to ``QCD'' results, we will understand predictions at NNLO QCD accuracy. For a generic observable $\Sigma$ in the additive approach we denote the prediction at this level of accuracy as $\Sigma_{\rm QCD+ EW}$. 

In the multiplicative approach one assumes complete factorisation of NLO QCD and NLO EW effects. This approach is presented in sec.~\ref{sec:multiplic} and is denoted as $\Sigma_{\rm QCD\times EW}$. The precise definition of the various quantities mentioned in the text is given in appendix \ref{sec:appendix} where an appropriate notation for the classification of the different contributions is introduced. Here, we just state  the most relevant definitions for the following discussion,
\noindent
\begin{eqnarray}
                                           & \Sigma_{\rm QCD} 
                                           &\equiv \Sigma_{\rm  LO~QCD} + \Sigma_{\rm NLO~QCD} + \Sigma_{\rm NNLO~QCD} \, , \nonumber \\
                                           &\Sigma_{\rm EW} &\equiv \Sigma_{\rm LO~EW} 
                                      + \Sigma_{\rm NLO~EW} + \SigmaSub\, , \nonumber \\
                                      &\Sigma_{\rm QCD+EW}  &\equiv \Sigma_{\rm  QCD} + \Sigma_{\rm EW}\, , \nonumber \\
                                      & \Sigma_{\rm QCD\times EW} &\equiv  \Sigma_{\rm QCD+EW}
                                                            +(K^{\rm NLO}_{\rm QCD}-1) \times \Sigma_{\rm NLO~EW}\, , \label{summarise:equations}
\end{eqnarray}
where $\Sigma$ denotes a generic observable in $t \bar t$ production and $K^{\rm NLO}_{\rm QCD}$ is the standard NLO/LO $K$-factor in QCD.
A variation of the multiplicative approach denoted as $ \Sigma_{\rm QCD^2 \times EW} $ will also be considered; it is defined similarly to $\Sigma_{\rm QCD\times EW}$  in eq.~\eqref{summarise:equations} but with NNLO/LO QCD $K$-factor.

As it has been discussed in ref.~\cite{Pagani:2016caq}, the usage of different PDF sets leads to a very different impact of the photon-induced contribution on $t\t$ distributions. While in the case of {\sc\small NNPDF3.0QED} the impact of photon-induced contributions is relatively large and with very large uncertainties, in the case of {\sc\small LUXQED} it is expected to be negligible. For this reason in the rest of this work we always show predictions with both PDF sets.

\subsection{Additive combination}\label{sec:main}

\begin{figure}[t]
\centering
\includegraphics[width=0.49\textwidth]{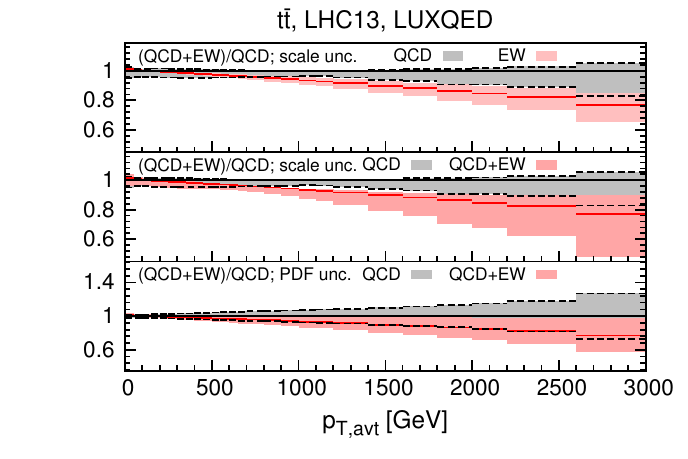}
\includegraphics[width=0.49\textwidth]{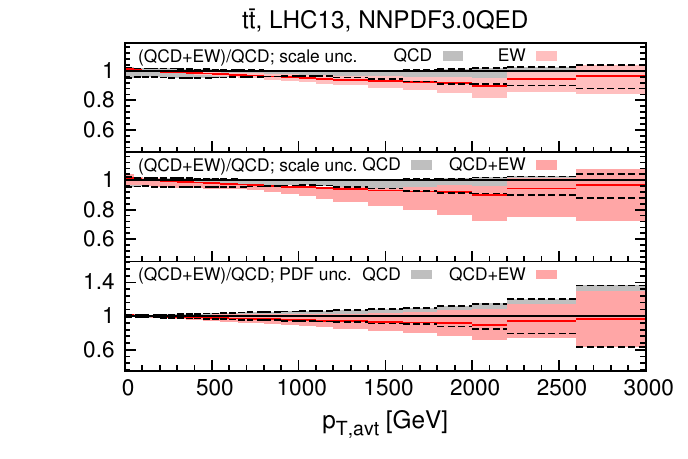}
\includegraphics[width=0.49\textwidth]{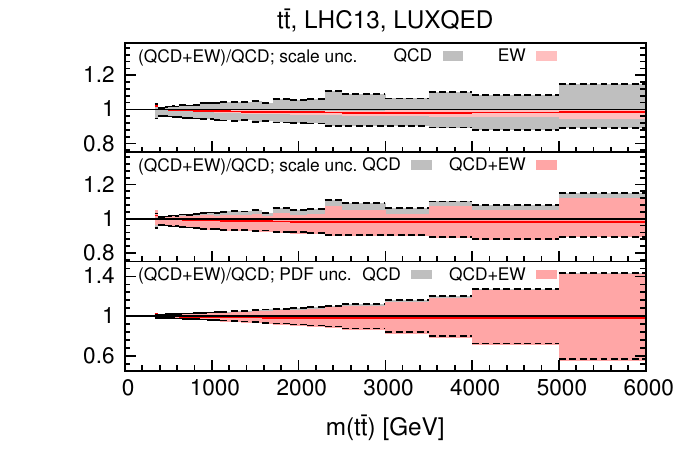}
\includegraphics[width=0.49\textwidth]{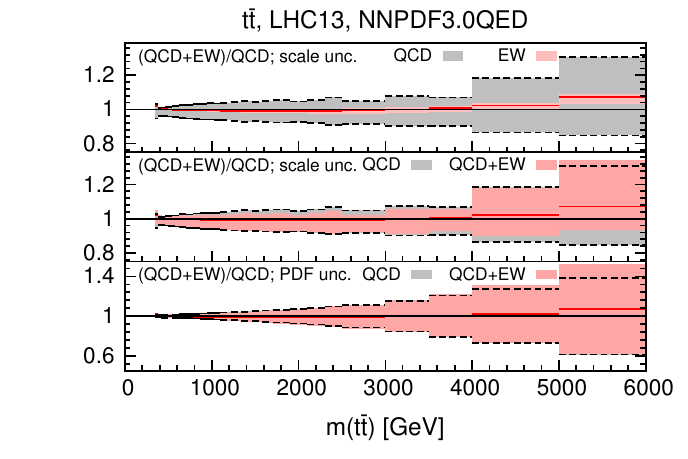}
\caption{Differential distributions for $\PTavt$ and  $m(t\t)$ at 13 TeV. Predictions are based on the additive approach. The format of the plots is described in the text.}
\label{fig:main_1}
\end{figure}

\begin{figure}[t]
\centering
\includegraphics[width=0.49\textwidth]{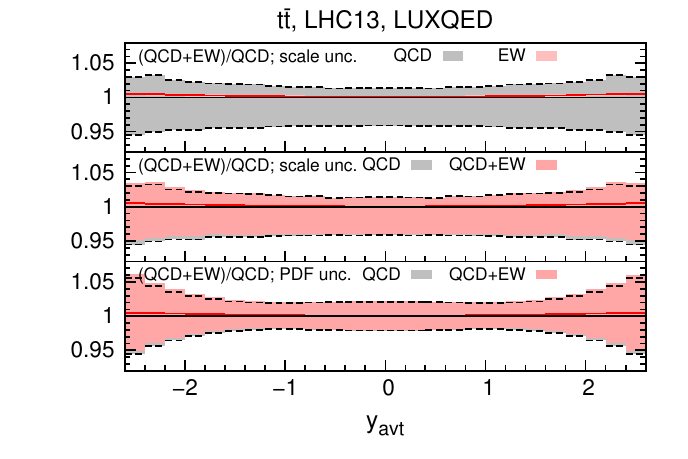}
\includegraphics[width=0.49\textwidth]{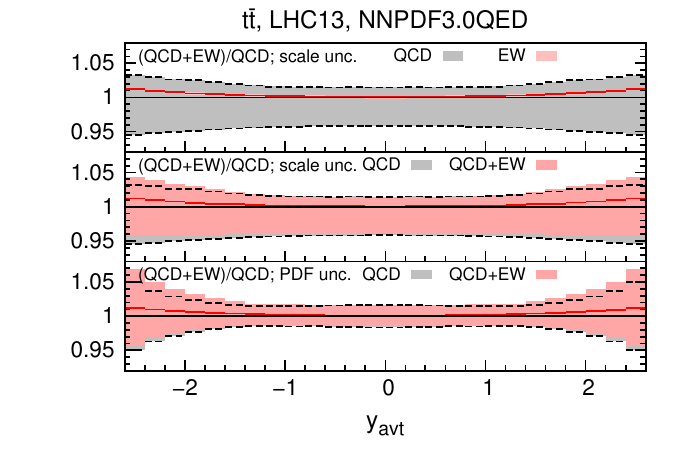}
\includegraphics[width=0.49\textwidth]{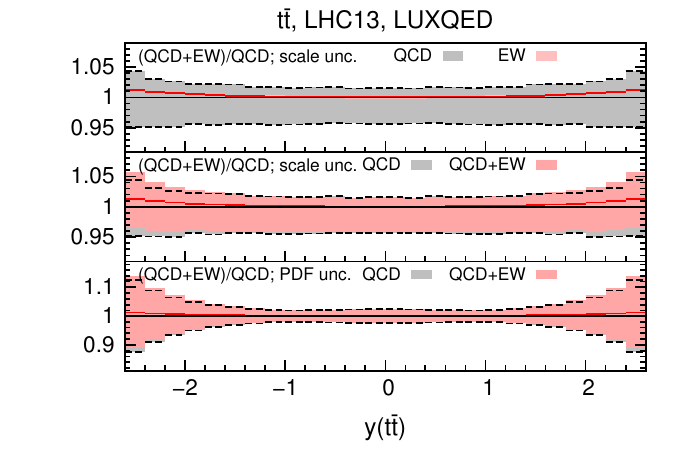}
\includegraphics[width=0.49\textwidth]{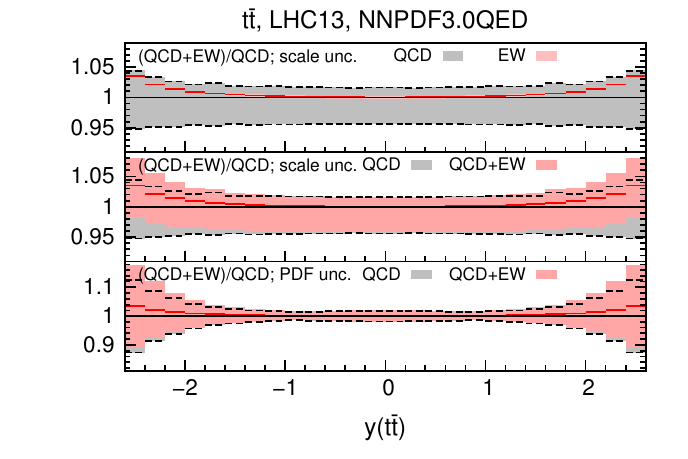}
\caption{As in fig.\ref{fig:main_1} but for the $\Yavt$ and $y(t\t)$ differential distributions.}
\label{fig:main_2}
\end{figure}

Distributions for $\PTavt$ and  $m(t\t)$ are shown in fig.~\ref{fig:main_1}, while the $\Yavt$ and $y(t\t)$ distributions are shown in fig.~\ref{fig:main_2}. 
The format of the plots for all distributions is as follows: for each observable, we show two plots side-by-side, with the same layout. The plot on the left-hand side shows predictions obtained using the {\sc\small LUXQED} set, while for the one on the right the {\sc\small NNPDF3.0QED} set is employed. Results at NNLO QCD accuracy are labelled  as ``QCD'' while the combination of NNLO QCD predictions and EW corrections in the additive approach are labelled as ``QCD+EW''. 

In each plot the three insets display ratios of different quantities~\footnote{It is actually in all cases the $\Sigma_{\rm QCD+ EW}/\Sigma_{\rm QCD}$ ratio, but the bands refer to three different quantities, as explained in the text.} over the central-scale QCD result (i.e., in the case of {\sc\small LUXQED}, the black line in the main panel of fig.~\ref{fig:pheno}). In the first inset we show the scale uncertainty due to EW corrections alone (red band), without QCD contributions ($\Sigma_{\rm EW}$ using the notation of Appendix \ref{sec:appendix}). This quantity can be compared to the scale uncertainty of the QCD prediction at NNLO accuracy (grey band). In the second inset we present the scale-uncertainty band (red) for the combined QCD+EW prediction. The grey band corresponds to the NNLO QCD scale-uncertainty band already shown in the first inset.
The third inset is equivalent to the second one, but it shows the PDF uncertainties.  We combine, for each one of the PDF members, the QCD prediction and the EW corrections into the QCD+EW result. The PDF uncertainty band of the QCD+EW prediction is shown in red while the grey band corresponds to the PDF uncertainty of the QCD prediction. For all insets, when the grey band is covered by the red one, its borders are displayed as black dashed lines.

As can be seen in figs.~\ref{fig:main_1} and \ref{fig:main_2}, the effect of EW corrections is, in general, within the NNLO QCD scale uncertainty. A notable exception is the case of the $\PTavt$ distribution with {\sc\small LUXQED}. In the tail of this distribution the effect of Sudakov logarithms is large and negative, of the order of -(10--20\%), and is not compensated by the photon-induced contribution. On the contrary,  in the case of {\sc\small NNPDF3.0QED}, photon-induced contributions mostly compensate the negative corrections due to Sudakov logarithms. As it has already been noted in ref.~\cite{Pagani:2016caq}, with this PDF set, the effect of photon-induced contributions is not negligible also for large values of $m(t\t)$, $\Yavt$ and $y(t\t)$.

As it can be seen in the first inset, in the large $\PTavt$ regime the scale dependence of the EW corrections alone is of the same size as, or even larger than, the scale variation at NNLO QCD accuracy. For this reason, as evident from the second inset, the scale uncertainty of the combined QCD+EW prediction is much larger than in the purely QCD case, both with the {\sc\small LUXQED} and {\sc\small NNPDF3.0QED} PDF sets. This feature is present only in the tail of the $\PTavt$ distribution.

The PDF uncertainties (third inset) for all distributions do not exhibit large differences between QCD and QCD+EW predictions, despite the fact that the photon-induced contribution in {\sc\small NNPDF3.0QED} is large and has very large PDF uncertainty (relative to {\sc\small LUXQED}).

\subsection{Multiplicative combination and comparison with the additive one}\label{sec:multiplic}

The additive approach $\Sigma_{\rm QCD+ EW}$ for combining QCD and EW corrections discussed in sec.~\ref{sec:main} is exact to the order at which the perturbative expansion of the production cross-section is truncated. An alternative possibility for combining QCD and EW corrections is what we already called the {\it multiplicative} approach, $\Sigma_{\rm QCD \times EW}$. This approach is designed to approximate the leading EW corrections at higher orders. In the case of $t\t$ production these are NNLO EW contributions of order ${\cal O}(\alphas^3 \alpha)$. 

The multiplicative approach is motivated by the fact that soft QCD and EW Sudakov logarithms factorise, with the latter typically leading to large negative corrections for boosted kinematics. Thus, when dominant NLO EW and NLO QCD corrections are at the same time induced by these two effects, the desired fixed order can be very well approximated via rescaling NLO EW corrections with NLO QCD $K$-factors. 
\footnote{The precise definitions of $\Sigma_{\rm QCD \times EW}$ is given in eq.~\eqref{eq:mult}  }
Otherwise, if one is in a kinematical regime for which the dominant NLO EW or NLO QCD corrections are of different origin (i.e. not Sudakov or soft), the difference between the multiplicative and additive approaches given by the term $\Sigma_{\rm mixed}$ in eq.~\eqref{mixed} can be considered as an indication of theory uncertainty in {\it that}\, kinematics. It must be stressed that the perturbative orders involved in the additive approach are included exactly also in the multiplicative approach; the only addition the multiplicative approach introduces on top of the additive one is the approximated ${\cal O}(\alphas^3 \alpha)$ contribution.

One of the advantages of the multiplicative approach is the stabilisation of scale dependence. As we saw in sec.~\ref{sec:main}, when QCD and EW corrections are combined in the additive approach, the scale dependence at large $\PTavt$ can exceed that of the NNLO QCD prediction. On the other hand, the large $\PTavt$ limit is precisely the kinematic regime where the multiplicative approach is a good approximation and can be trusted: at large $\PTavt$ the NLO EW and NLO QCD corrections are mainly induced by Sudakov logarithms and soft emissions, respectively, and as we just pointed out these two contributions factorise. The presence of large Sudakov logarithms in the NLO EW result at large $\PTavt$ is easy to see since for Born kinematics large $\PTt$ implies large $\PTtbar$ which, in turn, implies large $\hat{s}, \hat{t}$ and $\hat{u}$ Mandelstam variables. That NLO QCD corrections at large $\PTavt$ are mainly of soft origin can be shown with an explicit NLO calculation; by applying appropriate cuts on the jet, the top and/or the antitop one can easily see that the differential cross-section is dominated by kinematic configurations containing almost back-to-back hard top and antitop and a jet with small $p_{T}$. Plots demonstrating this can be found in footnote \ref{Rep}. 

In the following, for all observables $\Sigma$ considered in this work, we present predictions in the multiplicative approach denoted as $\Sigma_{\rm QCD\times EW}$.
As a further check of the stability of the multiplicative approach we display also the quantity $\Sigma_{\rm QCD^2 \times EW}$, whose precise definition can be found in appendix \ref{sec:appendix}. $\Sigma_{\rm QCD^2 \times EW}$ is defined analogously to $\Sigma_{\rm QCD \times EW}$, but by rescaling NLO EW corrections via NNLO QCD $K$-factors. By comparing $\Sigma_{\rm QCD \times EW}$ and $\Sigma_{\rm QCD^2 \times EW}$ one can further estimate the uncertainty due to mixed QCD-EW higher orders.

\begin{figure}[t]
\centering
\includegraphics[width=0.49\textwidth]{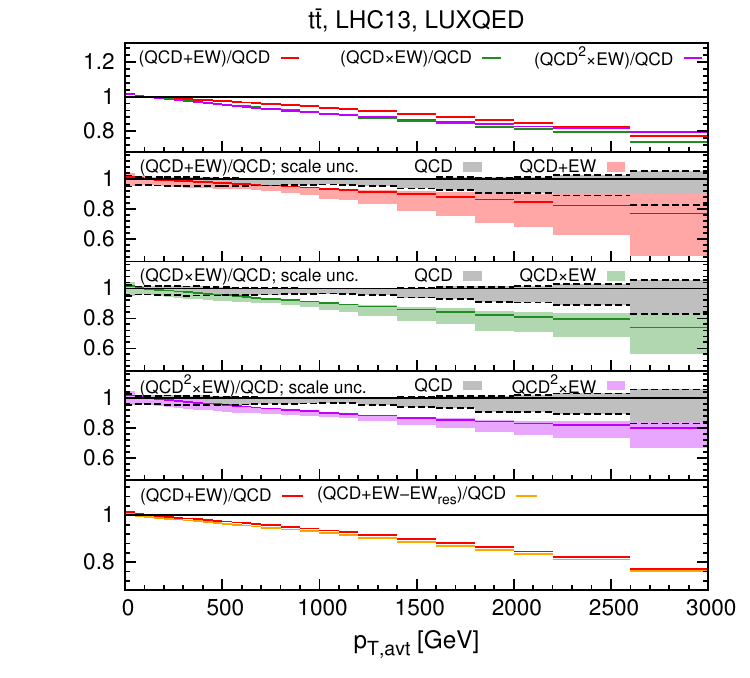}
\includegraphics[width=0.49\textwidth]{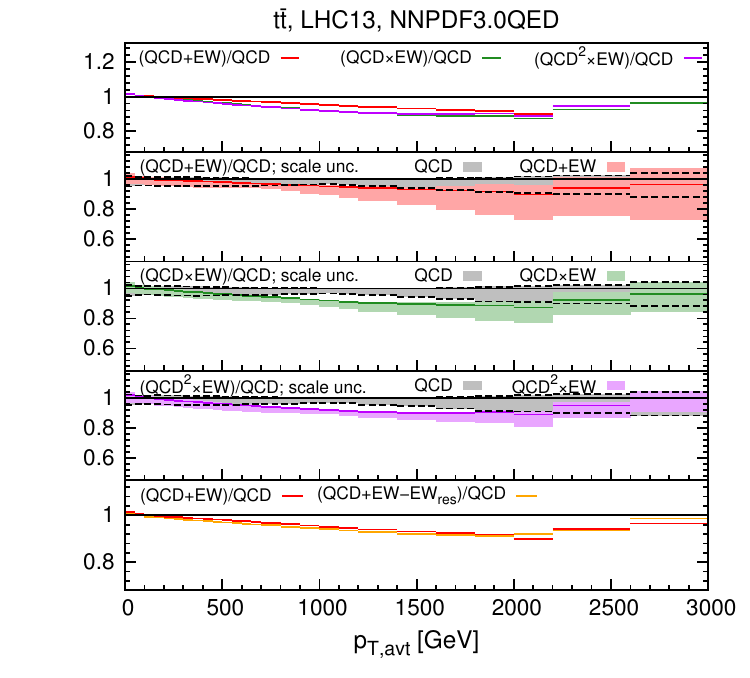}
\includegraphics[width=0.49\textwidth]{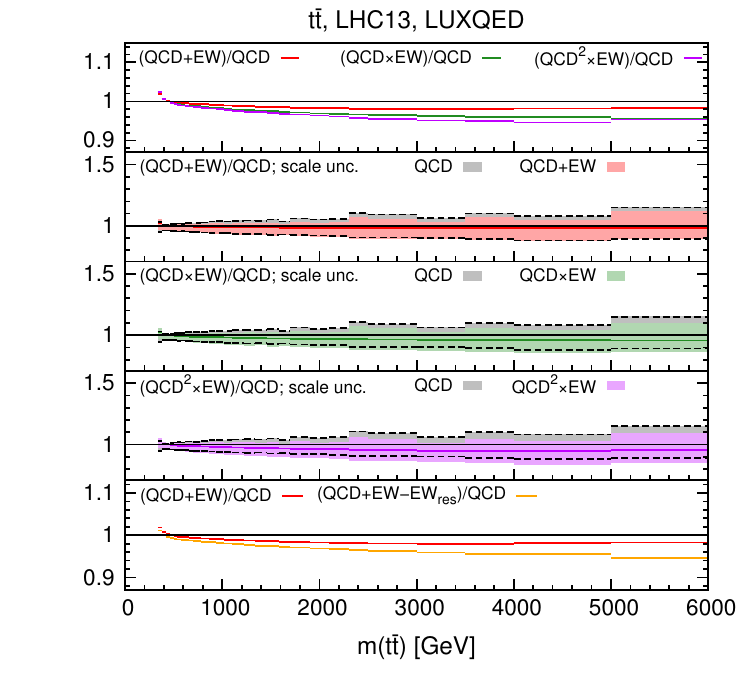}
\includegraphics[width=0.49\textwidth]{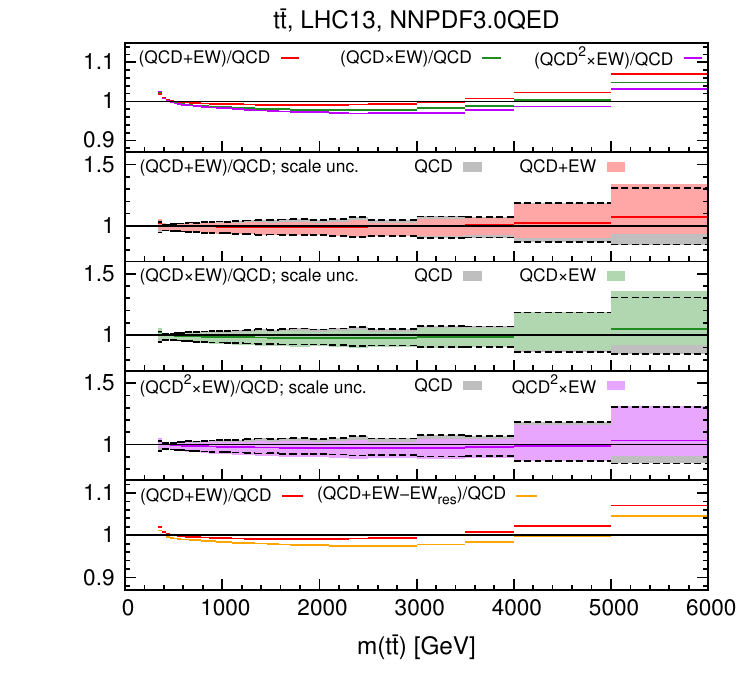}
\caption{Additive ($\Sigma_{\rm QCD+EW}$) versus multiplicative ($\Sigma_{\rm QCD\times EW}$) approach: $\PTavt$ and $\Mtt$ differential distributions at 13 TeV. The format of the plots is described in the text.}
\label{fig:mult_1}
\end{figure}
\begin{figure}[t]
\centering
\includegraphics[width=0.49\textwidth]{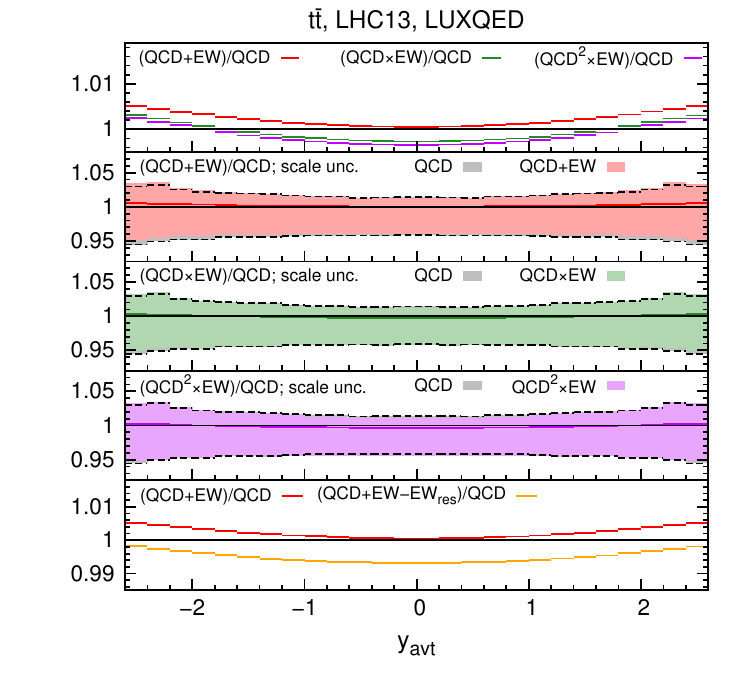}
\includegraphics[width=0.49\textwidth]{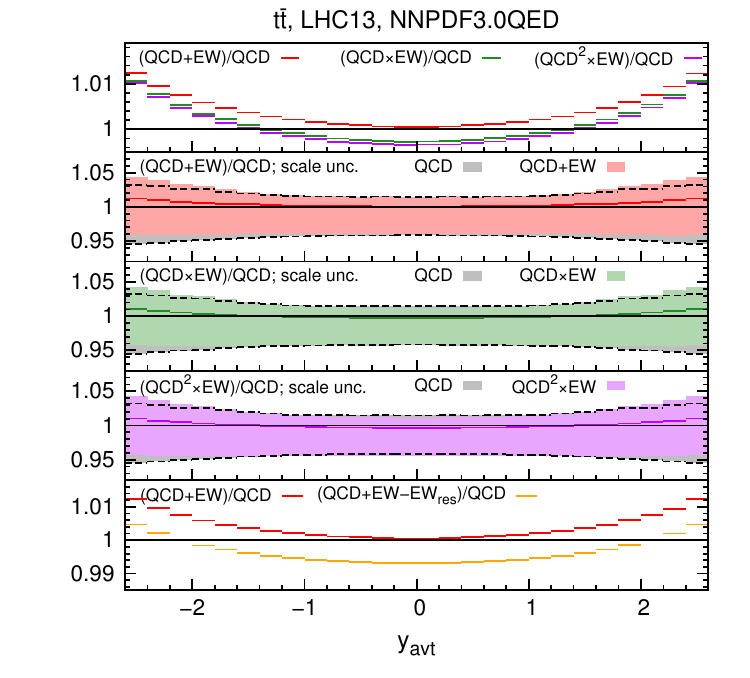}
\includegraphics[width=0.49\textwidth]{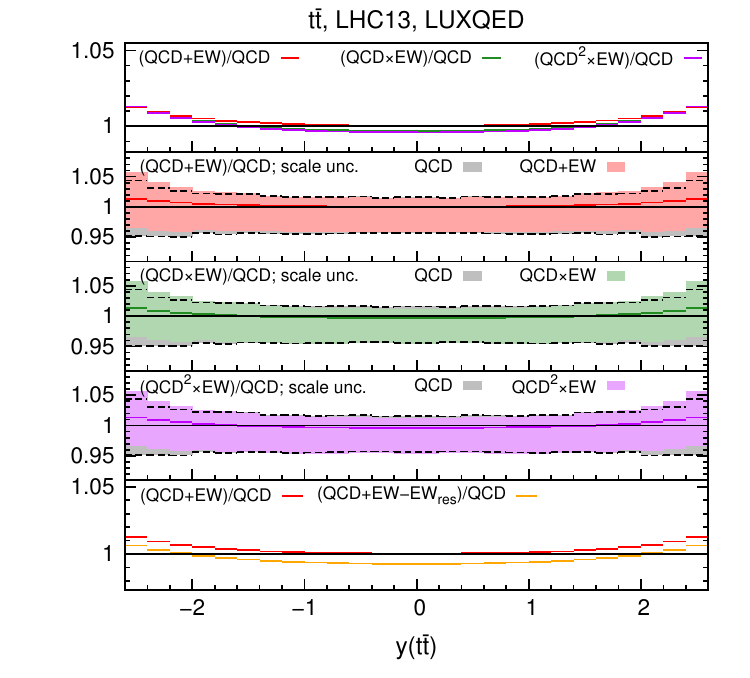}
\includegraphics[width=0.49\textwidth]{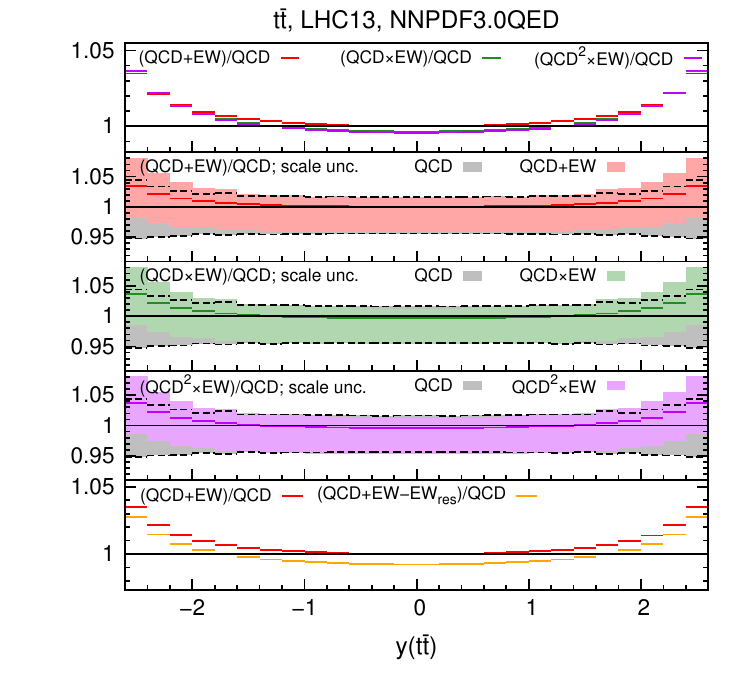}
\caption{As in fig.~\ref{fig:mult_1} but for the $\Yavt$ and $\Ytt$ differential distributions.}
\label{fig:mult_2}
\end{figure}

Figure~\ref{fig:mult_1} shows the $\PTavt$ and $\Mtt$ distributions, while fig.~\ref{fig:mult_2} refers to $\Yavt$ and $\Ytt$. As in sec.~\ref{sec:main}, the plots on the left are produced using the {\sc\small LUXQED} PDF set, while those on the right using the {\sc\small NNPDF3.0QED} PDF set. We next describe the format of the plots.

Each plot consists of five insets, which all show ratios of different quantities over the central value of $\Sigma_{\rm QCD}$. In the first inset we compare the central-scale results for the three alternative predictions: $\Sigma_{\rm QCD+EW}/\Sigma_{\rm QCD}$ (red line), $\Sigma_{\rm QCD\times EW}/\Sigma_{\rm QCD}$ (green line) and $\Sigma_{\rm QCD^2 \times EW}/\Sigma_{\rm QCD}$ (violet line). These quantities are further displayed in the second, third and fourth inset, respectively, where not only  the central value but also the scale dependence of the numerator is shown. In all cases we calculate the scale-uncertainty band as a scale-by-scale combination and subsequent variation in the 7-point approach. Scale variation bands have the same colour as the corresponding central-value line. For comparison we also display (grey band) the relative scale uncertainty of $\Sigma_{\rm QCD}$. Thus, the second inset is exactly the same as the second inset in the corresponding plots in sec.~\ref{sec:main}. The last inset shows a comparison of the ratio $\Sigma_{\rm QCD+EW}/\Sigma_{\rm QCD}$ including (red line) or not (orange line) the contribution $\SigmaSub$, where ``res'' stands for residual and denotes the fact that  $\SigmaSub$ are contributions to $\Sigma_{\rm EW}$ that are expected to be small, regardless of the PDF set used (see eq.~\eqref{blobend}).

As expected, the multiplicative approach shows much smaller dependence on the scale variation. This is particularly relevant for the tail of the $\PTavt$ distribution, where the scale uncertainty of $\Sigma_{\rm EW}$ alone is comparable in size with the one of $\Sigma_{\rm QCD}$; with this reduction of the scale uncertainty the $\Sigma_{\rm QCD\times EW}$ and $\Sigma_{\rm QCD}$ uncertainty bands do not overlap when {\sc\small LUXQED} is used. In the case of $\Mtt$ and $\Yavt$ distributions, the $\Sigma_{\rm QCD \times EW}$ central-value predictions are typically larger in absolute value than those of $\Sigma_{\rm QCD+EW}$, while they are all almost of the same size for the $\Ytt$ distribution. In the case of $\Yavt$ the difference between the additive and multiplicative approaches is completely negligible compared to their scale uncertainty.
Therefore, besides the kinematic region where Sudakov effects are the dominant contribution, the multiplicative and additive approaches are equivalent. Moreover, the difference between $\Sigma_{\rm QCD\times EW}$ and $\Sigma_{\rm QCD^2\times EW}$ is in general small; a sizeable difference between their scale dependences can be noted only in the tail of the $\PTavt$ distribution.  

For all the reasons mentioned above we believe that the multiplicative approach should be preferred over the additive one and, indeed, it has been used for the calculation of our best predictions in sec.~\ref{sec:pheno}. As can be seen from figs.~\ref{fig:mult_1} and \ref{fig:mult_2} and their threshold-zoomed-in versions in footnote \ref{Rep} the difference between $\Sigma_{\rm QCD + EW}$ and $\Sigma_{\rm QCD \times EW}$ for non-boosted kinematics is much smaller than the total theory uncertainty (scale+PDF) shown in fig.~\ref{fig:pheno}. Thus, the difference between the two approaches can be safely ignored in the estimation of the theory uncertainty. One should bear in mind that this conclusion depends on the choice of scale, which in our case, as explained in ref.~\cite{Czakon:2016dgf}, is based on the principle of fastest convergence. A different scale choice with larger $K$ factors would likely artificially enhance the difference between $\Sigma_{\rm QCD + EW}$ and $\Sigma_{\rm QCD \times EW}$.

In the last inset in figs.~\ref{fig:mult_1} and \ref{fig:mult_2} we compare the quantities $\Sigma_{\rm EW}$ and $\Sigma_{\rm EW}-\SigmaSub$, where the $\SigmaSub$ contribution is exactly included in both the additive and multiplicative approaches. As expected, one can see that the  $\SigmaSub$ contribution is typically flat and very small. The only exception is the $\Mtt$ distribution where a visible difference between the two curves ($\Sigma_{\rm EW}$ and $\Sigma_{\rm EW}-\SigmaSub$) is present, especially in the tail. The $\SigmaSub$ contribution includes the squared EW tree-level diagrams, the $\mathcal{O}(\alpha^2)$ contribution denoted as $\Sigma_{\rm LO,3}$ in \eqref{eq:blobs}, and the two subleading NLO corrections of respectively  $\mathcal{O}(\alpha_s \alpha^2)$ and $\mathcal{O}(\alpha^3)$, denoted as $\Sigma_{\rm NLO,3}$ and $\Sigma_{\rm NLO,4}$ in \eqref{eq:blobs}. While the $\mathcal{O}(\alpha^3)$ corrections are completely negligible, the $\mathcal{O}(\alpha^2)$ and $\mathcal{O}(\alpha_s \alpha^2)$ ones both lead to positive non-negligible contributions of similar size to the   $\Mtt$ distribution. Indeed, the $\mathcal{O}(\alpha^2)$ contribution involves $b \bar b \rightarrow t \bar t$ squared diagrams with a $W$ boson in the $t$-channel, which at large $\Mtt$ are not as much suppressed  as the contributions from the other initial states. Similarly, the $\mathcal{O}(\alpha_s \alpha^2)$ contributions contain QCD corrections to them, featuring the same beahaviour.  Relevant plots displaying individually all the   aforementioned contributions can be found at \ref{Rep}.

\section{Impact of the photon PDF}\label{sec:photonpdf}

\begin{figure}[t]
\centering
\includegraphics[width=0.49\textwidth]{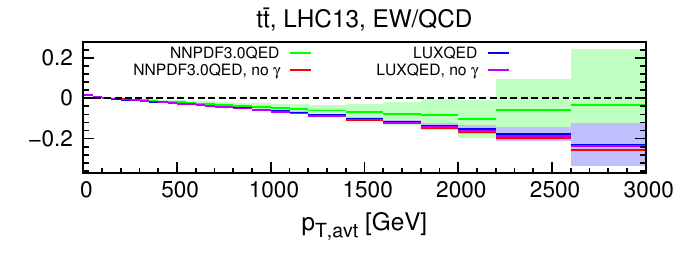}
\includegraphics[width=0.49\textwidth]{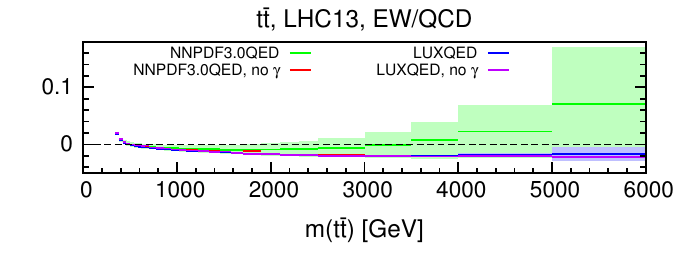}
\includegraphics[width=0.49\textwidth]{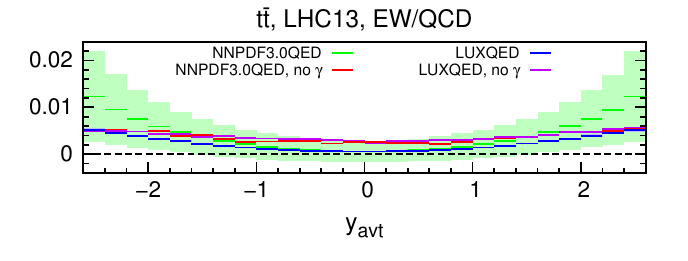}
\includegraphics[width=0.49\textwidth]{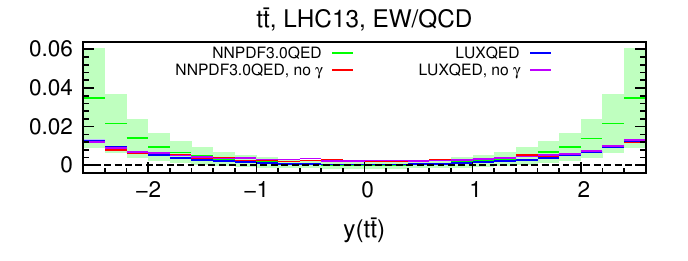}
\caption{Impact of the photon PDF on the $\PTavt$,  $\Mtt$, $\Yavt$ and $\Ytt$ differential distributions at 13 TeV. The format of the plots is described in the text.}
\label{fig:photonPDF}
\end{figure}

In this section we quantify the impact on $t\t$ differential distributions of the difference between the photon densities provided by the {\sc\small LUXQED} and  {\sc\small NNPDF3.0QED} PDF sets. In other words, we repeat the study performed in ref.~\cite{Pagani:2016caq} for these two PDF sets since they were not considered in that work. We compare the size of the electroweak corrections with and without the photon PDF for both PDF sets. In each plot of fig.~\ref{fig:photonPDF} we show the relative impact induced by the electroweak corrections (the ratio $\Sigma_{\rm EW}/\Sigma_{\rm QCD}$; see definitions in Appendix \ref{sec:appendix}) for four cases: {\sc\small NNPDF3.0QED} setting the photon PDF equal to zero (red) or not (green), and {\sc\small LUXQED} setting the photon PDF equal to zero (violet) or not (blue). For the cases including the photon PDF, we also show the PDF-uncertainty band of $\Sigma_{\rm EW}$. 

The impact of the photon-induced contribution can be evaluated via the difference between the green and red lines in the case of {\sc\small NNPDF3.0QED} and the difference between the blue and violet lines in the case of {\sc\small LUXQED}. As can be seen in fig.~\ref{fig:photonPDF}, the impact of the photon PDF on the $\PTavt$,  $\Mtt$, $\Yavt$ and $\Ytt$ distributions is negligible in the case of {\sc\small LUXQED}, while it is large and with very large uncertainties for the case of {\sc\small NNPDF3.0QED}, as already pointed out in  ref.~\cite{Pagani:2016caq}  for {\sc\small NNPDF2.3QED}. At very large $\PTavt$ and  $\Mtt$ also {\sc\small LUXQED} show a non-negligible relative PDF uncertainty, which is not induced by the photon but from the PDFs of the coloured partons at large $x$. We checked that a similar behaviour is exhibited also by NNPDF3.0QED when its photon PDF is set to zero.

\section{Contributions from heavy boson radiation}\label{sec:HBR}

In the calculation of EW corrections to QCD processes the inclusion of real emissions of massive gauge bosons (heavy boson radiation or HBR) is not mandatory since, due to the finite mass of the gauge bosons, real and virtual weak corrections are separately finite (albeit the virtual corrections are enhanced by large Sudakov logarithms).
Furthermore, such emissions are typically resolved in experimental analyses and are generally considered as a different process $t\t V(+X)$ with $V=H,W^{\pm},Z$. For these reasons, the results in sec.~\ref{sec:pheno} do not include HBR contributions.

It is, nonetheless, interesting to estimate the contribution of HBR to inclusive $t\t$ production. Our motivation is threefold: First, resolved or not, HBR is a legitimate contribution to the $t\t(+X)$ final state considered in this work. Secondly, it is clear that one cannot guarantee that HBR is resolved with 100\% efficiency. Therefore, it is mandatory to have a prior estimate for the size of the effect. Finally, we are unaware of prior works where the HBR contribution has been estimated in inclusive $t\t$ production. Recently, refs.~\cite{Frixione:2014qaa, Frixione:2015zaa} have provided estimates for HBR in the processes $t\t V(+X)$, with $V=H,Z,W$.

\begin{figure}[t]
\centering
\includegraphics[width=0.49\textwidth]{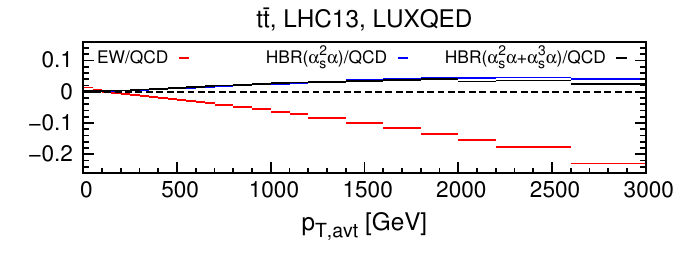}
\includegraphics[width=0.49\textwidth]{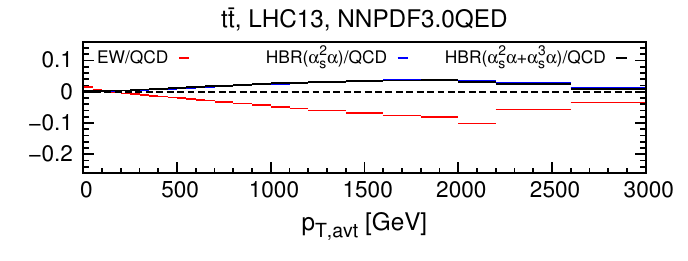}
\includegraphics[width=0.49\textwidth]{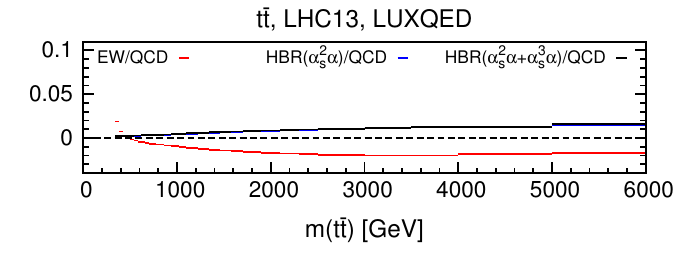}
\includegraphics[width=0.49\textwidth]{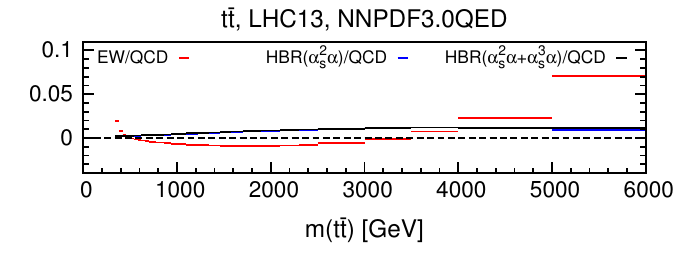}
\includegraphics[width=0.49\textwidth]{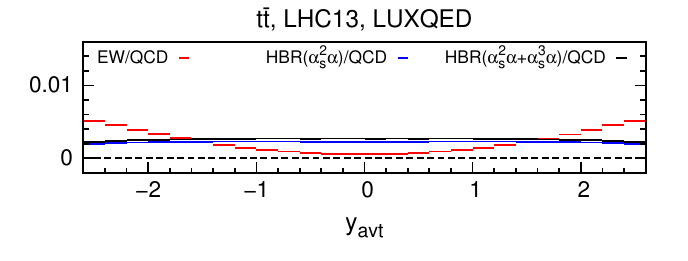}
\includegraphics[width=0.49\textwidth]{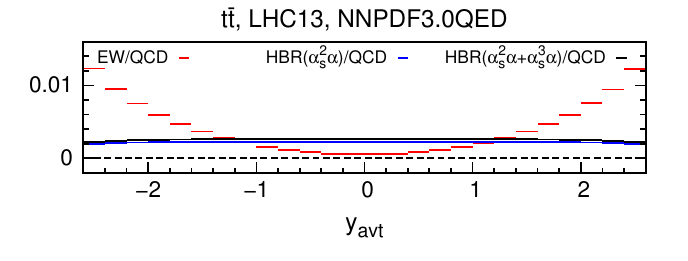}
\includegraphics[width=0.49\textwidth]{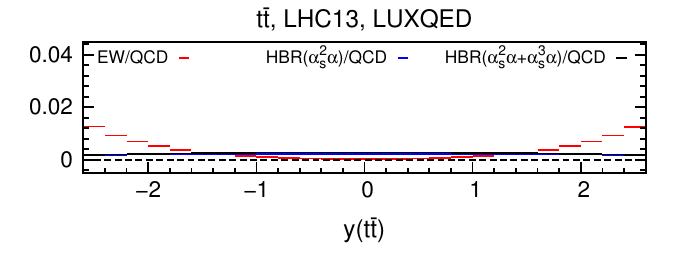}
\includegraphics[width=0.49\textwidth]{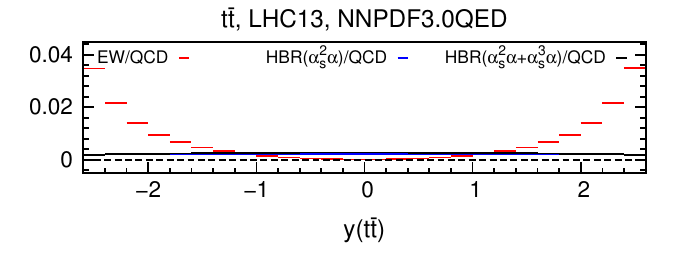}
\caption{Impact of heavy boson radiation (HBR) on the $\PTavt$,  $\Mtt$, $\Yavt$ and $\Ytt$ differential distributions at 13 TeV. The format of the plots is described in the text.}
\label{fig:HBR}
\end{figure}

We have investigated the impact of HBR on all four distributions considered in this work: $\PTavt$,  $\Mtt$, $\Yavt$ and $\Ytt$. Our results are shown in fig.~\ref{fig:HBR}, where we plot the effect of HBR on the central scale normalised to the QCD prediction. We show separately the LO HBR effect of order ${\cal O}( \alphas^2\alpha)$ as well as the NLO QCD HBR prediction which includes terms of order ${\cal O}(\alphas^3\alpha)$. As a reference we also show the EW corrections for $t \t$. 

In our calculations we include HBR due to $H,W$ and $Z$. We are fully inclusive in HBR, {\it i.e.}, no cuts on the emitted heavy bosons are applied. Clearly, any realistic experimental analysis will require an estimate of HBR subject to experimental cuts, but such an investigation would be well outside the scope of the present work.

From  fig.~\ref{fig:HBR} we conclude that the effect of HBR is generally much smaller than the EW corrections. In particular, higher-order QCD corrections to HBR are completely negligible, i.e. HBR is well described in LO for all the $t \t$ inclusive distributions and for the full kinematic ranges considered here. The absolute effect of HBR on the $\PTavt$ distribution is positive and small; it never exceeds 2-3\% (relative to the $t \t$ prediction at NNLO QCD accuracy) and is always much smaller than the EW correction. The only distribution where the HBR contribution is not negligible compared to the EW one is $\Mtt$ computed with {\sc\small LUXQED}. For this distribution the HBR correction is positive and only about half the absolute size of the (negative) EW correction. Still, the absolute size of the HBR, relative to the prediction at NNLO QCD accuracy, is within 1\% and so its phenomenological relevance is unclear. The impact of HBR on the two rapidity distributions is tiny, typically within 3 permil of the NNLO QCD prediction.

\section{Conclusions}

In this work we derive for the first time predictions for all main top-quark pair differential distributions 
\footnote{One distribution we do not consider is $p_{T,t\t}$ which is not known in NNLO QCD, and for which resummation is mandatory in order to have reliable predictions.}
with stable top quarks at the LHC at NNLO QCD accuracy and including the following EW corrections: the NLO EW effects of ${\cal O}(\alphas^2 \alpha)$, all subleading NLO terms of order ${\cal O}(\alphas \alpha^2)$ and ${\cal O}( \alpha^3)$ as well as the LO contributions of order ${\cal O}(\alphas \alpha)$ and ${\cal O}( \alpha^2)$.  

We present a detailed analysis of top-pair production at the LHC at 13 TeV and we find that the effect of EW corrections on differential distributions with stable top quarks is in general within the {\it current} total (scale+PDF) theory uncertainty. A notable exception is the $\PTavt$ distribution in the boosted regime where the effect of EW corrections is significant with respect to the current total theory error. We have checked that similar conclusions apply also for LHC at 8 TeV. All results derived in this work in the multiplicative approach, for both 8 and 13 TeV, are available in electronic form \ref{Rep} as well as with the ArXiv submission of this paper.

Providing phenomenological predictions for the LHC is only one of the motivations for the present study. In this work we also quantify the impact of the photon PDF on top-pair differential distributions and study the difference between the additive and multiplicative approaches for combining QCD and EW corrections. Moreover, we analyse the contribution from inclusive Heavy Boson Radiation on inclusive top-pair differential distributions.

In order to quantify the impact of the photon PDF, we use two recent PDF sets whose photon components are constructed within very different approaches. The first set, {\sc\small LUXQED}, is based on the {\sc\small PDF4LHC15} set \cite{Butterworth:2015oua} and adds to it a photon contribution that is derived from the structure function approach of ref.~\cite{Manohar:2016nzj}.  The second set, {\sc\small NNPDF3.0QED}, is based on the {\sc\small NNPDF3.0} family of PDFs and adds a photon component that is extracted from a fit to collider data.  {\sc\small NNPDF3.0QED} photon density has both a much larger central value and PDF uncertainty than those of  {\sc\small LUXQED}. On the other hand, the two sets are compatible within PDF errors and they both include QED effects in the DGLAP evolution on top of the usual NNLO QCD evolution.

We confirm the observations already made in ref.~\cite{Pagani:2016caq}, namely, the way the photon PDF is included impacts all differential distributions. The size of this impact is different  for the various distributions; the most significant impact can be observed in the $\PTavt$ distribution at moderate and large $p_T$ where the net effect from EW corrections based on {\sc\small NNPDF3.0QED} is rather small and with large PDF uncertainties, while using {\sc\small LUXQED} it is negative, with small PDF uncertainties and comparable to the size of the NNLO QCD scale error. The $\Mtt$ distribution displays even larger effects, but only at extremely high $\Mtt$. The $\Ytt$ distribution is also affected at large $\Ytt$ values.\footnote{As it has been lengthly motivated and discussed in ref.~\cite{Pagani:2016caq}, effects due to the photon PDF {\it \'{a} la} {\sc\small NNPDF} are even more pronounced at 8 TeV. } 

It seems to us that a consensus is emerging around the structure-function approach of ref.~\cite{Manohar:2016nzj}. Given its appealing predictiveness, this approach will likely be utilised in the future in other PDF sets. Therefore,  {\it at present}, it seems to us that as far as the photon PDF is concerned predictions based on the {\sc\small LUXQED} set should be preferred. 

Our best predictions in this work are based on the so-called multiplicative approach for combining QCD and EW corrections. We have also presented predictions based on the standard additive approach. In general, we find that the difference between the two approaches is small and well within the scale uncertainty band. The difference between the two approaches is more pronounced for the $\Mtt$ and $\PTavt$ distributions. Nevertheless, both approaches agree within the scale variation. The scale uncertainty is smaller within the multiplicative approach and, especially in the case of the $\PTavt$ distribution, does not overlap with the NNLO QCD uncertainty band. We stress that these features may be sensitive to the choice of factorisation and renormalisation scales.

Since we are unaware of a past study of Heavy Boson Radiation (i.e. $H,W^{\pm}$ and $Z$) in inclusive $t\t$ production, for completeness, we have also presented the impact of inclusive HBR on the inclusive top-pair differential spectrum. While it is often assumed that additional HBR emissions can be removed in the measurements, it is nevertheless instructive to consider the contribution of such final states. We find that, typically, the HBR contribution is negligible, except for the $\Mtt$ distribution, where it tends to partially offset the EW correction (when computed with {\sc\small LUXQED}).  We have also checked that NLO QCD corrections to the LO HBR result are negligible for all inclusive $t\t$ distributions considered by us.

\begin{acknowledgments}
We thank Stefano Frixione for his suggestions and interest at the early stage of this project. D.P., I.T. and M.Z. acknowledge also Fabio Maltoni for his strong encouragement and support  in pursuing this study.

The work of M.C. is supported in part by grants of the DFG and BMBF. The work of D.H. and A.M. is supported by the UK STFC grants ST/L002760/1 and ST/K004883/1. A.M. is also supported by the European Research Council Consolidator Grant ``NNLOforLHC2". The work of D.P is partially supported by the ERC grant 291377 ``LHCtheory: Theoretical predictions and analyses of LHC physics: advancing the precision frontier" and by the Alexander von Humboldt Foundation, in the framework of the Sofja Kovalevskaja Award Project ``Event Simulation for the Large Hadron Collider at High Precision''. The work of I.T. is supported by the F.R.S.-FNRS ``Fonds de la Recherche Scientifique'' (Belgium) and in part by the Belgian Federal Science Policy Office through the Interuniversity Attraction Pole P7/37. The work of M.Z. is supported by the European Union's Horizon 2020 research and innovation programme under the Marie Sklodovska-Curie grant agreement No 660171 and in part by the ILP LABEX (ANR-10-LABX-63), in turn supported by French state funds managed by the ANR within the ``Investissements d'Avenir'' programme under reference ANR-11-IDEX-0004-02.
\end{acknowledgments}
\\

\appendix\section{Notation}\label{sec:appendix}

In this appendix we specify how EW corrections and NNLO QCD results are combined in the additive and multiplicative approaches. The notation matches the one introduced in \cite{Frixione:2014qaa}. The phenomenology of $t\t$ production within the additive approach is presented in sec.~\ref{sec:main}. The multiplicative approach is studied in sec.~\ref{sec:multiplic} where it is also compared to the additive one.

A generic observable $\Sigma^{t\t}$ in the process $pp \to t\t(+X)$ can be expanded simultaneously in the QCD and EW coupling constants as:
\noindent
\begin{equation}
\Sigma^{t\t}(\alpha_s,\alpha) = \sum_{m+n\geq 2} \alpha_s^m \alpha^n \Sigma_{m+n,n}\, .
\end{equation}
\noindent
The LO ($m+n=2$), NLO ($m+n=3$) and NNLO ($m+n=4$) contributions read
\begin{align}
\Sigma^{t\t}_{\rm LO}(\alpha_s,\alpha) &= \alpha_s^2 \Sigma_{2,0} + \alpha_s \alpha \Sigma_{2,1} + \alpha^2 \Sigma_{2,2} \nonumber\\
 &\equiv \Sigma_{\rm LO,1} + \Sigma_{\rm LO,2} + \Sigma_{\rm LO,3}\, ,  \nonumber\\
\Sigma^{t\t}_{\rm NLO}(\alpha_s,\alpha) &= \alpha_s^3 \Sigma_{3,0} + \alpha_s^2 \alpha \Sigma_{3,1} + \alpha_s \alpha^2 \Sigma_{3,2} + \alpha^3 \Sigma_{3,3} \nonumber\\
 &\equiv \Sigma_{\rm NLO,1} + \Sigma_{\rm NLO,2} + \Sigma_{\rm NLO,3} + \Sigma_{\rm NLO,4}\, ,  \nonumber\\
\Sigma^{t\t}_{\rm NNLO}(\alpha_s,\alpha) &= \alpha_s^4 \Sigma_{4,0} + \alpha_s^3 \alpha \Sigma_{4,1} + \alpha_s^2 \alpha^2 \Sigma_{4,2} + \alphas \alpha^3 \Sigma_{4,3} +  \alpha^4 \Sigma_{4,4} \nonumber\\
 &\equiv \Sigma_{\rm NNLO,1} + \Sigma_{\rm NNLO,2} + \Sigma_{\rm NNLO,3} + \Sigma_{\rm NNLO,4} + \Sigma_{\rm NNLO,5} \;.
\label{eq:blobs}
\end{align}
\noindent
In order to simplify the notation, we further define the following purely QCD quantities
\noindent
\begin{align}
\Sigma_{\rm LO~QCD} &\equiv \Sigma_{\rm LO,1} \,, 
                                      &\Sigma_{\rm NLO~QCD} 
                                      &\equiv  \Sigma_{\rm NLO,1}\,, \label{blobbegin} \\
\Sigma_{\rm NNLO~QCD} &\equiv  \Sigma_{\rm NNLO,1}\, ,  
                                           & \Sigma_{\rm QCD} 
                                           &\equiv \Sigma_{\rm  LO~QCD} + \Sigma_{\rm NLO~QCD} + \Sigma_{\rm NNLO~QCD}
\end{align}
and those involving EW corrections
\begin{align}
\Sigma_{\rm LO~EW} &\equiv \Sigma_{\rm LO,2}  \,,
                                    &\SigmaSub &\equiv \Sigma_{\rm LO,3} 
                                    +  \Sigma_{\rm NLO,3} + \Sigma_{\rm NLO,4} \, ,\\
\Sigma_{\rm NLO~EW} &\equiv  \Sigma_{\rm NLO,2}  \,,
                                      &\Sigma_{\rm EW} &\equiv \Sigma_{\rm LO~EW} 
                                      + \Sigma_{\rm NLO~EW} + \SigmaSub\,. 
\label{blobend}
\end{align}
\noindent

Throughout this work with the term ``EW corrections'' we refer to the quantity $\Sigma_{\rm EW}$, while the term ``NLO EW corrections'' will only refer to $\Sigma_{\rm NLO~EW}$. In the additive approach, which is presented in section \ref{sec:main}, QCD and electroweak corrections are combined through the linear combination
\begin{equation} 
\Sigma_{\rm QCD+EW}  \equiv \Sigma_{\rm  QCD} + \Sigma_{\rm EW}\,.
\label{QCD+EW}
\end{equation}

The so called ``multiplicative approach'', which has been discussed in sec.~\ref{sec:main}, is precisely defined in the following. The purpose of the multiplicative approach is to estimate the size of $\Sigma_{\rm NNLO,2}$, which for convenience we rename $\Sigma_{\rm mixed}$ and assuming  complete factorisation of NLO QCD and NLO EW effects we estimate as
\begin{equation}
 \Sigma_{\rm mixed}\equiv \Sigma_{\rm NNLO,2} \approx \frac{\Sigma_{\rm NLO~QCD}\times \Sigma_{\rm NLO~EW}}{ \Sigma_{\rm LO~QCD}}\, . 
 \label{mixed}
 \end{equation}
  In the regime where NLO QCD corrections are dominated by soft interactions and NLO EW by Sudakov logarithms, eq.~\eqref{mixed} is a very good approximation, since the two effects factorise and are dominant. In other regimes $\Sigma_{\rm mixed}$ can be used as an estimate of the leading missing mixed QCD--EW higher orders. The advantage of the inclusion of $\Sigma_{\rm mixed}$ is the stabilisation of the scale dependence of the term $\Sigma_{\rm NLO ~EW}$, which in $t\t$ production has almost
\footnote{We say ``almost'' because this order receives also QCD corrections to the $\Sigma_{\rm LO ~EW}$ contributions from  the $g\gamma$ and $b\bar b$ initial states. Besides these effects $\Sigma_{\rm NLO ~EW}(\mu_2)=\Sigma_{\rm NLO ~EW}(\mu_1)\frac{\Sigma_{\rm LO~QCD}(\mu_2)}{\Sigma_{\rm {LO~QCD}(\mu_1)}}$.}
 the same functional form of $\Sigma_{\rm LO ~QCD}$. To this end we define the multiplicative approach as 
\noindent
\begin{align}
\Sigma_{\rm QCD\times EW} &\equiv  K^{\rm NLO}_{\rm EW} ~(\Sigma_{\rm LO~QCD} 
                                                             + \Sigma_{\rm NLO~QCD})+\Sigma_{\rm LO~EW}
                                                             + \Sigma_{\rm NNLO~QCD} 
                                                             + \SigmaSub \\
                                               & = K^{\rm NLO}_{\rm QCD} ~ (\Sigma_{\rm LO~QCD} 
                                                             + \Sigma_{\rm NLO~EW})
                                                             +\Sigma_{\rm LO~EW}
                                                             + \Sigma_{\rm NNLO~QCD} 
                                                             + \SigmaSub \\
                                               & = \Sigma_{\rm QCD}+K^{\rm NLO}_{\rm QCD} ~ \Sigma_{\rm NLO~EW} 
                                                             + \Sigma_{\rm LO~EW} 
                                                             + \SigmaSub\\
                                               & = \Sigma_{\rm QCD+EW}
                                                            +(K^{\rm NLO}_{\rm QCD}-1) \times \Sigma_{\rm NLO~EW}\\
                                                & \approx \Sigma_{\rm QCD+EW}
                                                            + \Sigma_{\rm mixed}\,,                                                  
\label{eq:mult}
\end{align}
where we used the standard $K$-factors 
\begin{equation}
K^{\rm NLO}_{\rm QCD}\equiv\frac{\Sigma_{\rm LO~QCD} 
+ \Sigma_{\rm NLO~QCD}}{\Sigma_{\rm LO~QCD}}\,, 
    \qquad K^{\rm NLO}_{\rm EW}\equiv\frac{\Sigma_{\rm LO~QCD} 
+ \Sigma_{\rm NLO~EW}}{\Sigma_{\rm LO~QCD}} \,.
\label{eq:Ks1}
\end{equation}

In order to test the stability of the multiplicative approach under even higher mixed QCD-EW orders, we combine NNLO QCD corrections and NLO EW corrections in order to estimate, besides the $\Sigma_{\rm mixed}$ term, also NNNLO contributions of order $\alpha_s^4 \alpha$. For this purpose we define the quantity
\noindent
\begin{align}
\Sigma_{\rm QCD^2\times EW} &\equiv K^{\rm NLO}_{\rm EW} ~ \Sigma_{\rm QCD}
                                                               +\Sigma_{\rm LO~EW}+ \SigmaSub\\
                                                   & = K^{\rm NNLO}_{\rm QCD} ~ (\Sigma_{\rm LO~QCD} 
                                                               + \Sigma_{\rm NLO~EW})+\Sigma_{\rm LO~EW} 
                                                               + \SigmaSub\\
                                                   & = \Sigma_{\rm QCD}+K^{\rm NNLO}_{\rm QCD} ~ \Sigma_{\rm NLO~EW} 
                                                               + \Sigma_{\rm LO~EW} 
                                                               + \SigmaSub\\
                                                   & = \Sigma_{\rm QCD+EW}
                                                               +(K^{\rm NNLO}_{\rm QCD}-1) \times \Sigma_{\rm NLO~EW}\,,
\label{eq:mult2}
\end{align}
where we introduced the  $K$-factor 
\begin{equation}
K^{\rm NNLO}_{\rm QCD}\equiv\frac{\Sigma_{\rm QCD} }{\Sigma_{\rm LO~QCD}}\,. 
\label{eq:Ks2} 
\end{equation}

Finally, we briefly describe how the dependence on the photon PDF enters the different perturbative orders. At LO and NLO accuracy, all contributions, with the exception of $\Sigma_{\rm LO~QCD}$ and $\Sigma_{\rm NLO~QCD}$, depend on the photon PDF. The dominant photon-induced process is $g \gamma  \to t\t$, which contributes to $\Sigma_{\rm LO~EW} $ and,  via QCD corrections to this order,  to $\Sigma_{\rm NLO~EW}$. In addition, $\Sigma_{\rm NLO~EW}$, but also $\Sigma_{\rm NLO,3}$ and $ \Sigma_{\rm NLO,4}$,  receive contributions from the $q\gamma\to t\t q$ and $\bar{q}\gamma\to t\t \bar{q}$ processes. Moreover, in the case of $\Sigma_{\rm LO,3}$ and $ \Sigma_{\rm NLO,4}$, also the $\gamma \gamma$ initial state contributes. As already discussed in ref.~\cite{Pagani:2016caq}, almost all of the photon-induced contribution arises form $\Sigma_{\rm LO~EW} $. In this work, at variance with ref.~\cite{Pagani:2016caq}, we also include the term $\SigmaSub$ in our calculations. However, since the size of $\SigmaSub$ is in general small, the previous argument still applies. The numerical impact of $\SigmaSub$ is discussed in sec.~\ref{sec:multiplic}.

Given the structure of the photon-induced contributions described before, it is also important to note that, with {\sc\small LUXQED}, the multiplicative approach is a better approximation of $\Sigma_{\rm mixed}$ than in the case of {\sc\small NNPDF3.0QED}. Indeed, the order $\Sigma_{\rm NLO~EW}$ contains also terms that can be seen as ``QCD corrections'' to the $g\gamma$  contributions in $\Sigma_{\rm LO,2}$ (negligible only with the {\sc\small LUXQED}), but are not taken into account in the multiplicative approach.

\bibliographystyle{JHEP}
\bibliography{}

\end{document}